\documentclass[twocolumn,pre,twoside,showpacs,showkeys,preprintnumbers,floatfix]{revtex4}

\usepackage{amsmath}
\usepackage{amssymb}
\usepackage{graphicx}
\usepackage{dcolumn}
\usepackage{bm}

\def\d{\mathrm{d}}
\def\epsilon{\varepsilon}
\def\enorm#1{\left\|{#1}\right\|_2}
\def\snorm#1{\left\|{#1}\right\|_\infty}
\def\theta{\vartheta}
\def\m#1{\mathrm{#1}}
\def\Int#1#2#3{\int\limits_{#1}\!\mathrm{d}^{#2}{#3}\;}
\def\Ref#1{$(\ref{#1})$}
\def\rho{\varrho}
\def\mean#1{\overline{#1}}
\def\set#1{\underline{#1}}
\def\vec#1{\mathbf{#1}}

\begin{document}


\title{Surface properties of fluids of charged platelike colloids}

\author{Markus Bier}
\email{bier@fluids.mpi-stuttgart.mpg.de}

\author{Ludger Harnau}

\author{S. Dietrich}

\affiliation{
   Max-Planck-Institut f\"ur Metallforschung, 
   Heisenbergstr. 3, 70569 Stuttgart, Germany, \\
   and Institut f\"ur Theoretische und Angewandte Physik,
   Universit\"at Stuttgart, Pfaffenwaldring 57, 70569 Stuttgart, Germany
}

\date{July 12, 2006}

\begin{abstract}
Surface properties of mixtures of charged platelike colloids and salt in contact with a charged
planar wall are studied within density functional theory. The particles are modeled by hard 
cuboids with their edges constrained to be parallel to the Cartesian axes corresponding to 
the Zwanzig model and the charges of the particles are concentrated in their centers. The density 
functional applied is an extension of a recently introduced functional for charged platelike colloids. 
Analytically and numerically calculated bulk and surface phase diagrams exhibit first-order wetting 
for sufficiently small macroion charges and isotropic bulk order as well as first-order drying 
for sufficiently large macroion charges and nematic bulk order. The asymptotic wetting and drying 
behavior is investigated by means of effective interface potentials which turn out
to be asymptotically the same as for a suitable neutral system governed by isotropic nonretarded 
dispersion forces. Wetting and drying points as well as predrying lines and the corresponding critical 
points have been located numerically. A crossover from monotonic to non-monotonic electrostatic potential 
profiles upon varying the surface charge density has been observed. Due to the presence of both the Coulomb 
interactions and the hard-core repulsions, the surface potential and the surface charge do 
not vanish simultaneously, i.e., the point of zero charge and the isoelectric point of the
surface do not coincide.
\end{abstract}



\maketitle


\section{Introduction}
Clay particles are examples of charged colloidal plates with diameters in the range from nanometer
to micrometer. Besides their technological importance in applications 
such as agriculture (as soil), construction (in concrete), filling (in cosmetics, 
rubber, plastic, etc.), coating (of paper), and oil-drilling (as rheological fluids 
for oil recovery), clay suspensions play an important role in basic research
because they exhibit a complex phase behavior in terms of packing and orientation of 
the platelike particles \cite{Mour95,Gabr96,Mour98,Brow99,Levi00,Kooi00,Beek03,Beek04a,
Wijn05,Ruzi06,Harn06}. Examples include sol-gel transitions, flocculation, aging, and even 
liquid crystalline phases depending on numerous parameters such as particle size, 
shape, charge, polydispersity, ionic strength, solvent, temperature, and external 
fields. The challenge is to understand the relation between these parameters 
and the phase behavior in order to control the properties of such fluids. 

Most studies in this direction have been devoted to \emph{bulk} properties.
Some experimental progress has been made due to advances in the synthesis of well 
defined and experimentally convenient model systems \cite{Brow98,Kooi98}. For a 
recent review of these investigations see, e.g., Ref. \cite{Davi05}. Computer 
simulations for platelike particles in bulk systems have been performed in recent 
years \cite{Veer92,Dijk97,Meye01a,Meye01b,Odri04,Beek04b,John05}, facing particular 
challenges as compared with systems consisting of rodlike particles. Most theoretical work 
has been concentrated on the
investigation of models of hard platelets \cite{Fors78,Cues99,Wens01,Rowa02,Harn02a,Wens04a}
and anisotropically charged discs \cite{Rose84,Rose85,Rowa00,Harn01,Harn02b,Triz02,Agra04,
Cost05,Li05} in spatially homogeneous configurations.

In contrast to the meanwhile well studied case of bulk suspensions of platelike particles, 
only a few theoretical investigations of such systems with \emph{spatial inhomogeneities} such as 
interfaces, surfaces, or due to external fields have been undertaken \cite{Moor92,Harn02c,
Harn02d,Bier04,Wens04b,Harn05,Bier05} because of the technical difficulties implied by reduced 
translational symmetry. On the other hand, understanding the influence of surfaces is of particular 
importance for applications because, e.g., omnipresent walls have a strong influence on this sort 
of material and electrodes are a common means to manipulate fluids of charged particles. A first 
important theoretical step for understanding such surface properties requires to determine the 
corresponding surface phase diagrams. Whereas surface phase transitions for simple fluid models 
have been investigated for decades \cite{Genn85,Sull86,Diet88,Schi90}, no such studies are 
available for complex fluids involving charges \emph{and} anisotropically shaped particles.

Here we present the investigation of a density functional theory for a fluid of charged
platelike colloids and salt in the presence of a charged hard wall. Due to the expected spatial
inhomogeneities of the fluid close to the surface, density functional theory \cite{Evan79} is 
the method of choice for addressing these issues \cite{Evan92}. Recently, we have proposed a 
density functional for fluids of charged platelike colloids which also takes long-ranged Coulomb 
interactions into account \cite{Bier05}. Here a straightforward extension of that functional is presented
in order to capture the influence of a charged hard wall. 

As introduced in Ref. \cite{Bier05}, the fluid is modeled by a mixture of colloidal rectangular plates and 
cubelike salt ions with the platelet orientation restricted to three mutually perpendicular 
directions (Zwanzig model \cite{Zwan63}). The charges are concentrated in the centers of the 
particles. The bulk phase diagram of this model exhibits one isotropic and one nematic phase separated 
by a first-order phase transition \cite{Bier05}. The salt concentration at coexistence of 
the isotropic and the nematic phase is larger for the isotropic than for the nematic state \cite{Bier05}.
This phenomenon is well-known for membrane equilibria \cite{Donn11,Donn24} where it is commonly 
referred to as the Donnan effect \cite{Adam73}. By bringing an isotropic and a coexisting nematic 
bulk system into spatial contact, an electric double layer forms at the free interface. For
negatively charged platelets, a negatively charged layer forms on the nematic side of the free
interface whereas a positively charged layer is found on the isotropic side \cite{Bier05}. The 
corresponding electrostatic potential profile changes from one side of the interface to the other 
by the so-called Donnan potential \cite{Adam73}. 

The present analysis is devoted to the investigation of the model fluid studied in Ref. \cite{Bier05} 
in contact with a hard wall which acquired charges upon releasing counterions into the fluid. Section 
\ref{sec:generalformalism} presents the model and extends the functional used in Ref. 
\cite{Bier05} in order to incorporate a wall potential. In Sec. \ref{sec:phasediagram}, the 
resulting bulk and surface phase diagrams are discussed. Further details concerning the wetting (drying) 
behavior of a charged wall in contact with an isotropic (nematic) bulk system are given in Sec. 
\ref{sec:wetting} (\ref{sec:drying}). Electrostatic potential profiles and the electrostatic 
surface potential are investigated in Sec. \ref{sec:electrostaticpotential}. Section \ref{sec:summary} 
summarizes the results.


\section{\label{sec:generalformalism}General formalism}

\subsection{Definitions}

We consider a fluid of charged platelike colloids and salt in three-dimensional space confined 
by a single planar structureless charged hard wall. 

The fluid model used here is the same as in Ref. \cite{Bier05}. It is a ternary mixture of 
charged hard square cuboids (see 
\begin{figure}[!t]
   \includegraphics[scale=0.8]{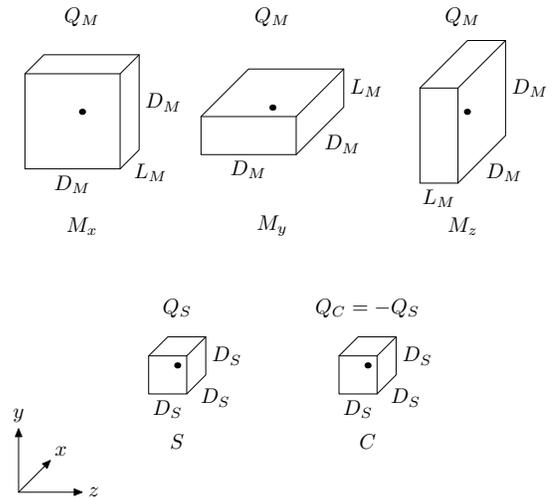}
   \caption{\label{fig:geo}Macroions $M$ are square cuboids of size $D_M \times D_M \times 
           L_M$, $D_M \not= L_M$ with charge $Q_M$ whereas salt ions $S$ and counterions 
           $C$ are cubes of side length $D_S$ with charges $Q_S$ and $-Q_S$, respectively.
           The pointlike charges ($\bullet$) are localized at the centers of the cuboids. The macroions can 
           adopt three possible orientations $M_x$, $M_y$, and $M_z$ corresponding to the 
           $L_M$-edges being parallel to the $x$-, $y$-, and $z$-axis, respectively.}
\end{figure}
Fig.~\ref{fig:geo}) with their edges parallel to the Cartesian axes 
(Zwanzig model \cite{Zwan63}) and dissolved in a dielectric solvent (e.g., water) with dielectric 
constant $\epsilon$. The solvent is treated as a continuum. For simplicity, the charges are 
fixed, monodisperse, and concentrated at the centers of the particles. The particles of the 
first component represent the macroions $M$ of size $D_M \times D_M \times L_M$, $D_M \not= L_M$, 
and charge $Q_M$. Within the Zwanzig approximation, macroions can take three different orientations, 
denoted as $M_x$, $M_y$, or $M_z$ corresponding to whether the $L_M$-edges are parallel to the $x$-, 
$y$-, or $z$-axis, respectively (see Fig.~\ref{fig:geo}). The second component consists of salt ions 
$S$ modeled as cubes of side length $D_S$ and charge $Q_S$ (see Fig.~\ref{fig:geo}). The salt ion charge 
$Q_S$ is chosen to have the same sign as the macroion charge $Q_M$ if the latter does not vanish. Finally, 
the third component consists of counterions $C$ guaranteeing overall charge 
neutrality. They are also described by cubes with the same side length $D_S$ but 
opposite charge $Q_C := -Q_S$ (see Fig.~\ref{fig:geo}). Due to the choice of the same sign
for $Q_M$ and $Q_S$, ions of component $C$ are counterions in the usual sense for both the
macroions $M$ and the salt ions $S$.

The wall, which confines the fluid from one side, is modeled as impenetrable for the fluid 
particles. Moreover, it carries a fixed homogeneously distributed surface charge of areal density $\sigma$ 
which, depending on the sign of $\sigma$, can be assumed to have been generated by releasing an appropriate
amount of ions of either component $S$ or component $C$ into the fluid.

$\rho_i(\vec{r}), i\in\{M_x,M_y,M_z, S, C\}$ denotes the number density at
point $\vec{r}$ of the centers of macroions with orientation $M_{x,y,z}$, salt ions, and counterions, 
respectively. Note that the position $\vec{r} \in V \subseteq\mathbb{R}^3$ with $V$ denoting the
system volume is a \emph{continuous} variable in contrast to the orientation of macroions, which 
varies within a \emph{discrete} set. We introduce the abbreviation
$\set{\rho}(\vec{r}) := (\rho_{M_x}(\vec{r}),\dots,\rho_C(\vec{r}))$. 

The system under consideration is coupled to two particle reservoirs: One supplies neutralized 
macroions (chemical formula $C_kM, k := \frac{Q_M}{Q_S} \geq 0$) and the 
other neutral salt (chemical formula $CS$); $\mu_{C_kM}$ and $\mu_{CS}$ denote the corresponding 
chemical potentials. These molecules dissociate upon entering the system:
\begin{eqnarray}
   C_kM & \longrightarrow & kC^{Q_C} + M^{Q_M}, 
   \nonumber\\
   CS   & \longrightarrow & C^{Q_C} + S^{Q_S}.
   \label{eq:dissoc}
\end{eqnarray} 
These equilibrium chemical reactions lead to the following relations between the reservoir chemical
potentials ($\mu_{C_kM}$ and $\mu_{CS}$) and the particle chemical potentials 
($\mu_i, i\in\{M_x,M_y,M_z, S, C\}, \mu_{M_x} = \mu_{M_y} = \mu_{M_z}$): 
\begin{eqnarray}
   \mu_{C_kM} & = & k\mu_C + \mu_{M_{x,y,z}} 
   \nonumber\\
   \mu_{CS} & = & \mu_C + \mu_S. 
   \label{eq:murel}
\end{eqnarray}

\subsection{Density-functional theory}

Configurations of the system are characterized by the set of number density profiles $\set{\rho}$. 
The equilibrium states minimize the grand canonical density functional \cite{Evan79,Evan92,Units}
\begin{eqnarray}
   \Omega[\set{\rho}] & = &  
   \sum_i\Int{V}{3}{r}\rho_i(\vec{r})\big(\ln(\rho_i(\vec{r})) - 1 - \mu^*_i + V_i(\vec{r})\big)
   \nonumber\\
   & & + F^\m{ex}[\set{\rho}] + U^\m{self}(\sigma),
   \label{eq:df1}
\end{eqnarray}
where $V_i$ is the wall (or substrate) potential exerted on particles of type $i$, $F^\m{ex}$ is the 
free energy in excess over the ideal gas contribution, and $U^\m{self}(\sigma)$ is the self energy of the 
charged wall. The latter is \emph{independent} of $\set{\rho}$ and hence merely shifts $\Omega$ by a 
constant value. It has been introduced for later convenience. Finally, the reduced particle chemical 
potentials $\mu^*_i := \mu_i - \ln\big(\Lambda_i^3\big)$ with the thermal de Broglie wavelength 
$\Lambda_i$ ($\Lambda_{M_x} = \Lambda_{M_y} = \Lambda_{M_z}$) for particles of type $i$ have been 
introduced. With the reduced reservoir chemical potentials
\begin{eqnarray}
   \mu_{C_kM}^* & := & \mu_{C_kM} - \Big(\ln\big(\Lambda_{M_{x,y,z}}^3\big) + 
                                        k \ln\big(\Lambda_C^3\big)\Big)
   \nonumber\\
   \mu_{CS}^* & := & \mu_{CS} - \Big(\ln\big(\Lambda_S^3\big) + 
                                        \ln\big(\Lambda_C^3\big)\Big),
   \label{eq:resmustar}
\end{eqnarray}   
Eq. \Ref{eq:murel} takes the form
\begin{eqnarray}
   \mu^*_{C_kM} & = & k\mu^*_C + \mu^*_{M_{x,y,z}} 
   \nonumber\\
   \mu^*_{CS} & = & \mu^*_C + \mu^*_S. 
   \label{eq:mustarrel}
\end{eqnarray}
For given reservoir chemical potentials $\mu^*_{C_kM}$ and $\mu^*_{CS}$, the particle chemical potentials
$\mu^*_i$ are fixed by Eq. \Ref{eq:mustarrel} and the constraint of global charge neutrality:
\begin{equation}
   \Int{V}{3}{r}\sum_iQ_i\rho_i(\vec{r}) + |A|\sigma= 0
   \label{eq:gcn}
\end{equation}
where $|A|$ is the area of the charged surface.

The Euler-Lagrange equations of the minimization problem read
\begin{equation}
   \frac{\delta \Omega}{\delta\rho_i(\vec{r})} = 
   \ln(\rho_i(\vec{r})) - \mu^*_i + V_i(\vec{r}) - c_i(\vec{r}) = 0
   \label{eq:ele}
\end{equation}
with the one-particle direct correlation function
\begin{equation}
   c_i(\vec{r}) := -\frac{\delta F^\m{ex}}{\delta\rho_i(\vec{r})}.
   \label{eq:1dcf}
\end{equation}
Equation \Ref{eq:ele} has to be solved under suitable boundary conditions in accordance with the bulk 
states for given reservoir chemical potentials $\mu_{C_kM}^*$ and $\mu_{CS}^*$. These bulk phases have 
already been determined in Ref. \cite{Bier05}.

\vspace*{20pt}

\subsection{\label{sec:Fex}Excess free energy}

As introduced in Ref. \cite{Bier05}, the interactions between charges are approximated by
\begin{equation}
   U^\m{c}_{ij}(\vec{r}-\vec{r'}) := 
   \frac{Q_iQ_j}{\snorm{\vec{r}-\vec{r'}}},
   \label{eq:Uc}
\end{equation}
where the usual Euclidean norm $\enorm{\vec{r}} = \sqrt{x^2+y^2+z^2}$ is replaced by the supremum 
norm $\snorm{\vec{r}} = \max(|x|,|y|,|z|)$ which is the adapted norm in the context of a Zwanzig 
model for cuboids. 

Employing a Debye charging process, $F^\m{ex}$ can be expressed as \cite{Bier05}
\begin{equation}
   F^\m{ex} = F^\m{ex,h} + F^\m{ex,c}_\m{el} + F^\m{ex,c}_\m{corr}.
   \label{eq:Fex2}
\end{equation}

The first term in Eq. \Ref{eq:Fex2} is the fundamental measure functional introduced by Cuesta and 
Mart\'\i nez-Rat\'on \cite{Cues97a,Cues97b}:
\begin{equation}
   F^\m{ex,h}[\set{\rho}] := \Int{V}{3}{r}\Phi(\set{n}(\vec{r}))
   \label{eq:Fexh}
\end{equation}
with the weighted densities
\begin{equation}
   n_\alpha(\vec{r}) = \sum_i\Int{V}{3}{r'} \omega_{\alpha,i}(\vec{r}-\vec{r'})\rho_i(\vec{r'})
   \label{eq:n}
\end{equation}
for $\alpha\in\{0,1x,1y,1z,2x,2y,2z,3\}$ and the excess free energy density
\begin{eqnarray}
   \Phi(\set{n}) & = &
   -n_0\ln(1-n_3) + \sum_{q\in\{x,y,z\}}\!\frac{n_{1q}n_{2q}}{1-n_3} + 
   \frac{n_{2x}n_{2y}n_{2z}}{(1-n_3)^2}.
   \nonumber\\
   & &
   \label{eq:Phi}
\end{eqnarray}
It describes the free energy due to the hard-core interactions $U^\m{h}_{ij}(\vec{r}-\vec{r'})$ between 
two particles of type $i$ and $j$ at positions $\vec{r}$ and $\vec{r'}$, respectively. 
$U^\m{h}_{ij}(\vec{r}-\vec{r'})$ is infinite if the hard cores of these particles overlap and zero otherwise.

The electrostatic part
\begin{equation}
   F^\m{ex,c}_\m{el}[\set{\rho}] 
   := 
   \frac{1}{2}\sum_{ij}\Int{V}{3}{r}\Int{V}{3}{r'}\rho_i(\vec{r})\rho_j(\vec{r'})
   U^\m{c}_{ij}(\vec{r}-\vec{r'})
   \label{eq:Fexcel}
\end{equation}
accounts for the particle-particle Coulomb interaction within the random phase approximation
(RPA) whereas the correlation part
\begin{widetext}
\begin{eqnarray}
   F^\m{ex,c}_\m{corr}[\set{\rho}] 
   & := & 
   \frac{1}{2}\sum_{ij}\Int{V}{3}{r}\Int{V}{3}{r'}\rho_i(\vec{r})\rho_j(\vec{r'})
   U^\m{c}_{ij}(\vec{r}-\vec{r'})
   \big(
   \exp(-U^\m{h}_{ij}(\vec{r}-\vec{r'}))-1
   \nonumber\\
   & & 
   + \exp(-U^\m{h}_{ij}(\vec{r}-\vec{r'}))
   G_{ij}(\kappa(\vec{r},\vec{r'};[\set{\rho}]),\snorm{\vec{r}-\vec{r'}})
   \big)
   \label{eq:Fexccorr}
\end{eqnarray}
\end{widetext}
with the function
\begin{equation}
   G_{ij}(\kappa,s) 
   :=
   -\int\limits_0^1\!\d\eta\; \min\big(1, U^\m{c}_{ij}(s)\eta\exp(-\sqrt{\eta}\kappa s)\big)
   \label{eq:Gdef}
\end{equation}
and the screening factor 
\begin{equation}
   \kappa(\vec{r},\vec{r'};[\set{\rho}]) := 
   \frac{1}{2}(\tilde\kappa(\vec{r};[\set{\rho}]) + \tilde\kappa(\vec{r'};[\set{\rho}]))
   \label{eq:kappa}
\end{equation}
using
\begin{equation}
   \tilde\kappa(\vec{r};[\set{\rho}]) := 
   \sqrt{4\pi Q_S^2(\rho_S(\vec{r}) + \rho_C(\vec{r}))}
   \label{eq:kappatilde}
\end{equation}
represents corrections beyond RPA.

Since $G_{ij}(\kappa,s\rightarrow\infty) = -12Q_iQ_j\kappa^{-4}s^{-5}$ (see Ref. \cite{Bier05}), 
the contribution $v_{ij}(s) := U^\m{c}_{ij}(s)G_{ij}(\kappa,s)$ 
in Eq. \Ref{eq:Fexccorr} \emph{beyond} RPA due to the Coulomb interactions 
has the \emph{same} $s^{-6}$ asymptotics and sign as the contribution \emph{within} RPA of nonretarded 
dispersion forces. The amplitudes of nonretarded van der Waals forces for typical Hamaker constants of 
clay (e.g., $6.6\ k_BT$ for $\m{Al_2O_3}$ \cite{Ackl96}) are two orders of magnitude smaller than those 
of $v_{ij}(s)$ for $|Q_M| = e$. Moreover, for long distances ($s\rightarrow\infty$) dispersion 
forces cross over to an $s^{-7}$ asymptotics due to retardation \cite{Diet88}. Dispersion forces are 
negligible in this retardation limit compared to $v_{ij}(s)$. For these reasons we have not included 
dispersion forces into our model.

\subsection{Wall potential}

Since the thermodynamic limit of globally charge neutral systems of hard particles with
Coulomb interactions exists \cite{Lebo69,Lieb72}, i.e., the bulk free energy density 
exists and it depends neither on the shape nor on the nature of the boundaries of the system volume $V$, 
the system volumes $V(L) := A(L) \times [0^-,L]$ of size $2L \times 2L \times L$ are 
considered in the limit $L \rightarrow \infty$. Here $A(L)$ is a square in the 
$x$-$y$ plane of side length $2L$, i.e., $|A(L)| = 4L^2$, with (lateral) periodic boundary conditions. 
Furthermore, the charged hard surface is chosen as the intersection of $V(L)$ with the
plane $z=0$. The notation $0^-$ in the above definition of $V(L)$ is 
used in order to emphasize the fact that the charged wall at $z=0$ belongs \emph{entirely} to
the system volume.

The wall potential $V_i$ for particles of type $i$ can be decomposed as 
$V^\m{h}_i + V^\m{c}_i$ with the hard wall contribution
\begin{equation}
   V^\m{h}_i(\vec{r}) := 
   \left\{\begin{array}{ll}
      \infty & , z \leq\frac{S^z_i}{2} \\[5pt]
      0      & , z > \frac{S^z_i}{2}
   \end{array}\right.,
   \label{eq:Vh}
\end{equation}
where $S^z_i$ is the $z$-extension of particles of type $i$, and the contribution due
to the surface charge
\begin{equation}
   V^\m{c}_i(\vec{r}) :=    
   \Int{V}{3}{r'}
   \frac{Q_i\sigma\delta(z')}{\snorm{\vec{r}-\vec{r'}}}.
\end{equation}
Similarly, the wall self energy is given by
\begin{equation}
   U^\m{self}(\sigma) :=    
   \frac{1}{2}\Int{V}{3}{r}\Int{V}{3}{r'}
   \frac{\sigma\delta(z)\sigma\delta(z')}{\snorm{\vec{r}-\vec{r'}}}.
\end{equation}
Introducing the local charge density
\begin{equation}
   \rho^Q(\vec{r}) := \sum_iQ_i\rho_i(\vec{r}) + \sigma\delta(z)
   \label{eq:rhoc}
\end{equation}
and the electrostatic potential
\begin{equation}
   \psi(\vec{r}) := \Int{V}{3}{r'} \frac{\rho^Q(\vec{r'})}{\snorm{\vec{r}-\vec{r'}}}
   \label{eq:psi1}
\end{equation}
one readily finds
\begin{eqnarray}
   & & 
   F^\m{ex,c}_\m{el} + \sum_i\Int{V}{3}{r}\rho_i(\vec{r})V_i^\m{c}(\vec{r}) + U^\m{self}(\sigma) 
   \nonumber\\
   & = &
   \frac{1}{2}\Int{V}{3}{r}\Int{V}{3}{r'}
   \frac{\rho^Q(\vec{r})\rho^Q(\vec{r'})}{\snorm{\vec{r}-\vec{r'}}}
   \nonumber\\
   & = &
   \frac{1}{2}\Int{V}{3}{r}\rho^Q(\vec{r})\psi(\vec{r}).
   \label{eq:sumel}
\end{eqnarray}

According to Ref. \cite{Bier05}, the expression for the electrostatic potential reduces to 
\begin{equation}
   \psi(z) = -4\int\limits_{0^-_{\vphantom{|}}}^L\!\d z'\; \rho^Q(z')|z-z'|,
   \label{eq:psi3}
\end{equation}
due to the lateral translational symmetry. By differentiating twice one can show that $\psi$ 
fulfills the Poisson equation
\begin{equation}
   \psi'' = -8\rho^Q.
   \label{eq:poisson}
\end{equation}

Combined with Eqs. \Ref{eq:Fex2} and \Ref{eq:sumel}, the density functional in 
Eq. \Ref{eq:df1} takes the final form
\begin{eqnarray}
   \Omega[\set{\rho}] 
   & = & 
   |A|\sum_i\int\limits_{0^-_{\vphantom{|}}}^L\!\d z\; \rho_i(z)
   \big(
   \ln(\rho_i(z)) - 1 - \mu^*_i 
   \nonumber\\
   & &
   + V_i^\m{h}(z) + \frac{1}{2}Q_i\psi(z)\big)
   + \frac{1}{2}|A|\sigma\psi(0)
   \nonumber\\
   & &
   + F^\m{ex,h}[\set{\rho}] + F^\m{ex,c}_\m{corr}[\set{\rho}] 
   \label{eq:df3}
\end{eqnarray}
which has to be minimized under the constraint of global charge neutrality 
\begin{equation}
   \int\limits_{0^-_{\vphantom{|}}}^L\!\d z\; \rho^Q(z) = 0.
\end{equation}
The corresponding Euler-Lagrange equations read
\begin{equation}
   \ln(\rho_i(z)) - \mu^*_i + V_i^\m{h}(z) + Q_i\psi(z) 
   - c_i^\m{h}(z) - c^\m{c}_{\m{corr},i}(z) = 0.
   \label{eq:ele2}
\end{equation}

\subsection{\label{sec:asym}Asymptotic behavior}

Expanding the Euler-Lagrange equations (Eq.~\Ref{eq:ele2}) for $z\rightarrow\infty$ around the bulk values 
provides the number density deviations $\Delta\set{\rho}(z) := \set{\rho}(z)-\set{\rho}(\infty)$ with
\begin{eqnarray}
   \Delta\rho_i(z\rightarrow\infty) & = & \rho_i(\infty)\big(-\Delta V_i^\m{h}(z) - Q_i\Delta\psi(z) 
   \label{eq:eleasym1}\\
   & &
   \phantom{\rho_i(\infty)\big(} + \Delta c_i^\m{h}(z) + \Delta c^\m{c}_{\m{corr},i}(z)\big).
   \nonumber
\end{eqnarray}
The electrostatic potential difference $\Delta\psi$ depends on $\Delta\set{\rho}$ via the local charge 
density $\rho^Q$ (see Eqs.~\Ref{eq:rhoc} and \Ref{eq:poisson}). However, this dependence will not be used
for the derivation of, c.f., Eq.~\Ref{eq:eleasym2}. Inserting Eqs.~\Ref{eq:Fexh} and \Ref{eq:Fexccorr} into 
Eq.~\Ref{eq:1dcf} leads to
\begin{equation}
   \Delta c_i^\m{h}(z\rightarrow\infty) = -\sum_j\Delta\rho_j(z)A_{ij}
   \label{eq:deltach}
\end{equation}
and
\begin{equation}
   \Delta c^\m{c}_{\m{corr},i}(z\rightarrow\infty) = \sum_j\big(t_{ij}(z) - \Delta\rho_j(z)B_{ij}\big),
   \label{eq:deltacccorr}
\end{equation}
where $t_{ij}(z) := -24Q_i^2Q_j^2\rho_j(\infty)\tilde\kappa(\infty)^{-4}z^{-3}$. 
The spatially constant $5 \times 5$-matrices $A_{ij}$ and $B_{ij}$ depend only on model 
parameters (particle sizes and charges) and bulk quantities (see Appendix A). 
By defining $M_{ij} := \delta_{ij}/\rho_i(\infty) + A_{ij} + B_{ij}$ and since $V^\m{h}_i(z) = 0$ for 
$z > \frac{1}{2}\max_i(S_i^z)$ (Eq. \Ref{eq:Vh}), Eq.~\Ref{eq:eleasym1} is equivalent to
\begin{equation}
   \sum_j M_{ij}\Delta\rho_j(z\rightarrow\infty) = -Q_i\Delta\psi(z) + \sum_j t_{ij}(z).
   \label{eq:eleasym2}
\end{equation}

Firstly, Eq.~\Ref{eq:eleasym2} implies an asymptotic decay of $\Delta\rho_i(z)$ \emph{not faster} than proportional
to $z^{-3}$; otherwise $\Delta\psi(z\rightarrow) = Kz^{-3}$ with some spatially constant amplitude $K$ would 
fulfill the
relation $K + 24Q_i\tilde\kappa(\infty)^{-4}\sum_jQ_j^2\rho_j(\infty)=0$ \emph{simultaneously} for \emph{all}
$i$ which is impossible because the second term depends on $i$ whereas the first does not.
Moreover, it can be shown that the deviations $\Delta\rho_i(z)$ exhibit a purely algebraic decay in leading order.
Finally, one is led to the conclusion that the asymptotic Euler-Lagrange equations \Ref{eq:eleasym2} in conjunction
with the Poisson equation \Ref{eq:poisson} leads to the properties 
$\Delta\rho_i(z\rightarrow\infty) = \mathcal{O}(z^{-3})$, $\Delta\psi(z\rightarrow\infty) = \mathcal{O}(z^{-3})$, 
and $\rho^Q(z\rightarrow\infty) = \mathcal{O}(z^{-5})$. Note that on the right-hand side of Eq.~\Ref{eq:rhoc} for 
$z\rightarrow\infty$ the leading and the next to leading order contributions to the density profiles $\set{\rho}$ 
cancel. 

Interestingly, the asymptotic decay proportional to $z^{-3}$ of $\Delta\rho_i(z)$ within the current model 
for charged particles equals the corresponding one for systems governed by isotropic nonretarded van der Waals 
forces in Ref.~\cite{Diet91}. This behavior is in sharp contrast to the results obtained within multicomponent 
Poisson-Boltzmann theories which give rise to exponentially decaying density profiles \cite{Grah47}.


\section{\label{sec:phasediagram}Bulk and surface phase diagram}

Upon solving the Euler-Lagrange equations (Eq. \Ref{eq:ele2}) for mixtures of platelike
macroions and monovalent salt in contact with a charged hard wall, one finds the
equilibrium state of this system. Here the equilibrium states are calculated as function
of the chemical potential $\mu^*_{C_kM}$, the macroion charge $Q_M$, and the surface charge 
density $\sigma$. The remaining parameters of the model have been fixed as \cite{Units}
(see Fig.~\ref{fig:geo}) $D_M=20\ell_B\approx 14\,\m{nm}$, $L_M=\ell_B\approx 0.72\,\m{nm}$, 
$D_S=\ell_B\approx 0.72\,\m{nm}$, $|Q_S|=e$, and salt density $\rho_S = 0.1\,\m{mM}$ in the bulk.
As discussed in Ref. \cite{Bier05}, the current model yields reasonable results only in
the range $|Q_M| \lesssim e$, i.e., close to the point of zero charge of the colloidal platelets.

It turns out that the equilibrium states exhibit rotational symmetry with respect to the wall 
normal ($z$-axis), i.e., $\rho_{M_x} = \rho_{My}$. Therefore, the local equilibrium structure of 
the macroions is captured completely by the two densities $\rho_{M_x}$ and $\rho_{M_z}$. Equivalently, 
the (total) macroion density $\rho_M := 2\rho_{M_x} + \rho_{M_z}$ and the nematic order parameter along
the $z$-axis
\begin{equation}
   s_M := \frac{3}{2}\frac{\rho_{M_z}}{\rho_M} - \frac{1}{2} \in \left[-\frac{1}{2},1\right]
\end{equation}
may serve to identify various structures. The definition of $s_M$ coincides with the well-known
scalar liquid-crystal order parameter 
$S = \langle P_2(\cos\theta)\rangle = \frac{3}{2}\langle(\cos\theta)^2\rangle - \frac{1}{2}$ 
for the special case of a Zwanzig model, within which the only possible macroion orientations are parallel 
($M_z$, $\cos\theta=1$) and perpendicular ($M_{x,y}$, $\cos\theta=0$) to the $z$-axis. Therefore,
structures with $s_M = 0$ and $s_M > 0$ are called \emph{isotropic} and \emph{nematic}, respectively.

\begin{figure}[!t]
   \includegraphics{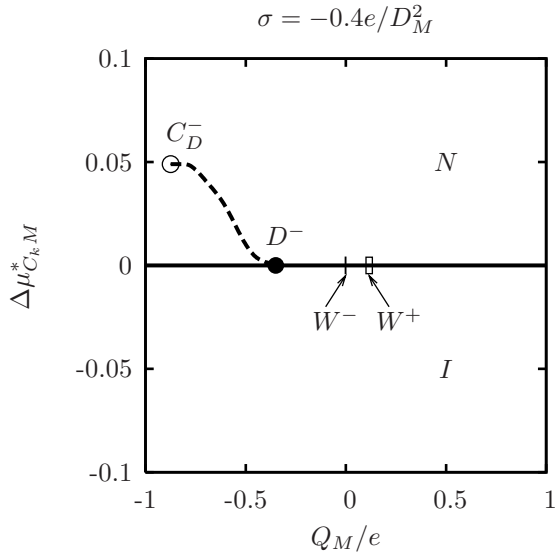}
   \caption{\label{fig:a}Bulk and surface phase diagram of mixtures \cite{Units} of platelike 
           macroions ($D_M=20\ell_B, L_M=\ell_B$) and monovalent salt ($D_S=\ell_B, |Q_S|=e$)
           for salt density $\rho_S=0.1\,\m{mM}$ in contact with a charged hard wall
           of surface charge density $\sigma = -0.4e/D_M^2$ in terms of macroion charge $Q_M$
           and chemical potential difference $\Delta\mu^*_{C_kM}$.
           The bulk equilibrium states for $\Delta\mu^*_{C_kM} < 0$ and $\Delta\mu^*_{C_kM} > 0$ 
           are isotropic ($I$) and nematic ($N$), respectively. 
           Isotropic-nematic bulk coexistence corresponds to $\Delta\mu^*_{C_kM} = 0$ 
           (solid line). For isotropic boundary conditions in the bulk ($z\rightarrow\infty$) at 
           isotropic-nematic coexistence ($\Delta\mu^*_{C_kM} = 0^-$), two first-order wetting transition
           points $W^-$ (at $Q_M^{W^-}\in[-2\times 10^{-3}e,0]$) and $W^+$ (at $Q_M^{W^+}\in[0.1e,0.132e]$) 
           have been found. The tolerance intervals are indicated by frames of corresponding widths.
           Complete wetting by the nematic phase occurs for $Q_M \in (Q_M^{W-},Q_M^{W^+})$
           upon approaching coexistence from the isotropic side. For nematic boundary 
           conditions in the bulk at isotropic-nematic coexistence ($\Delta\mu^*_{C_kM} = 0^+$), a 
           first-order drying transition point $D^-$ (at $Q_M^{D^-}=-0.35e$) has been found. The accompanying 
           predrying line (dashed line) terminates at a critical point $C_D^-$ (at $Q_M^{C_D^-} = -0.87e, 
           \Delta\mu^{*C_D^-}_{C_kM} = 0.049$). Complete drying by the isotropic phase occurs upon approaching
           coexistence from the nematic side for $Q_M < Q_M^{D^-}$. This implies that for 
           $Q_M^{D^-} < Q_M < Q_M^{W^-}$ or $Q_M^{W^-} < Q_M < e$ there is neither complete wetting 
           (by the nematic phase) nor complete drying (by the isotropic phase). A second drying transition 
           point $D^+$ appears (not shown) for $Q_M$ sufficiently large and $\sigma$ sufficiently small so that 
           there is reentrance of complete drying for large positive values of $Q_M$.}
\end{figure}
Figure~\ref{fig:a} displays the bulk and surface phase diagram for the surface charge density 
$\sigma=-0.4e/D_M^2$ in terms of the macroion charge $Q_M$ and the chemical potential 
difference $\Delta\mu^*_{C_kM}:=\mu^*_{C_kM}-\mu^{*IN}_{C_kM}$. The solid line 
($\Delta\mu^*_{C_kM}=0$) denotes the states of bulk coexistence between the isotropic ($I$) phase 
and the nematic ($N$) phase, corresponding to the chemical potential $\mu^{*IN}_{C_kM}$ at coexistence. 
The bulk equilibrium states for $\Delta\mu^*_{C_kM} < 0$ and $\Delta\mu^*_{C_kM} > 0$ are isotropic and 
nematic, respectively. 

At isotropic-nematic bulk coexistence with isotropic boundary conditions in the bulk
($\Delta\mu^*_{C_kM} = 0^-$), two first-order wetting transition points $W^-$ and $W^+$ have been found which 
could be located only within the intervals of the coexistence line indicated by frames of corresponding widths: 
Whereas $W^-$ ($Q_M^{W^-}\in[-2\times 10^{-3}e,0]$) is known rather precisely, there remains some uncertainty 
with respect to the location of $W^+$ ($Q_M^{W^+}\in[0.1e,0.132e]$). The corresponding prewetting lines are so
close to the coexistence line such that they could not be resolved numerically. 

Complete wetting by 
the nematic phase occurs for approaching isotropic-nematic bulk coexistence from the isotropic side for 
$Q_M \in (Q_M^{W^-},Q_M^{W^+})$. The phenomenon that only \emph{partial} wetting is found
for sufficiently large macroion charges $|Q_M|$ may be qualitatively understood as follows: The macroion 
number density profiles $\rho_{M_x}$, $\rho_{M_y}$, and $\rho_{M_z}$ close to the surface are influenced 
by the hard-core interactions --- which give rise to a preference of nematic order close to the wall ---, the 
macroion-substrate Coulomb interactions proportional to $|Q_M|$, and the macroion-macroion Coulomb 
repulsion proportional to $|Q_M|^2$. The latter dominates for large macroion charges $|Q_M|$ leading to 
a depression of the values of the macroion number densities near the wall which in turn prevents the 
growth of a nematic film. For small macroion charges $|Q_M|$, $\rho_{M_x}$, $\rho_{M_y}$, and $\rho_{M_z}$
near the surface are determined by the balance between the hard-core interactions and the 
macroion-substrate interactions which lead to complete wetting for attractive walls. 
Further details of the wetting behavior will be given in Sec. \ref{sec:wetting}.

In Fig.~\ref{fig:a}, a first-order drying transition point $D^-$ at $Q_M^{D^-}=-0.35e$ is found for 
isotropic-nematic bulk coexistence with nematic boundary conditions in the bulk ($\Delta\mu^*_{C_kM} = 0^+$). 
The first-order character of the drying transition at state point $D$ implies the 
existence of a predrying line (dashed line in Fig.~\ref{fig:a}), along which the excess adsorption of macroions
\begin{equation}
   \Gamma_M := \int\limits_0^\infty\!\d z\; (\rho_M(z)-\rho_M(\infty)),
   \label{eq:GammaM}
\end{equation}
which is proportional to the thickness of the emerging film, exhibits a finite discontinuity. The 
predrying line is expected to meet the isotropic-nematic bulk coexistence line tangentially \cite{Haug83}.
Our numerical data are consistent with this behavior but not definitely conclusive due to the numerical
difficulties arising from the large thickness of drying films very close to isotropic-nematic bulk 
coexistence. The predrying line terminates at a critical point $C_D^-$ located at
state point $(Q_M^{C_D^-} = -0.87e, \Delta\mu^{*C_D^-}_{C_kM} = 0.049)$. Complete drying by 
the isotropic phase occurs for $Q_M < Q_M^{D^-}$ upon approaching isotropic-nematic bulk coexistence from 
the nematic side. As for the wetting
scenario discussed above, the macroion-macroion repulsion will prevail over the macroion-surface
interactions for a sufficiently large macroion charges $|Q_M|$. Therefore, for large $|Q_M|$,
the formation of an isotropic film is initiated by the strongly depressed macroion number density close 
to the surface. Hence one expects complete drying for sufficiently large macroion charges $|Q_M|$. 
For a surface charge density $\sigma=-0.4e/D_M^2$ (see Fig.~\ref{fig:a}), \emph{partial} drying is found
for $Q_M \in (Q_M^{D^-},e]$, i.e., the expected second drying transition point $D^+$ is located
in the range $Q_M > e$. For smaller surface charge densities $\sigma$, indeed \emph{two} 
drying transition points have been found within the range $Q_M \in [-e,e]$. More details of the first-order 
drying transition and the corresponding predrying line can be found in Sec. \ref{sec:drying}.

Within the intervals $Q_M\in[Q_M^{D^-},Q_M^{W^-}]$ and $Q_M\in[Q_M^{W^+},Q_M^{D^+}]$ only partial wetting 
and drying occurs.
In order to validate the topology of the bulk and surface phase diagram in Fig.~\ref{fig:a}, we 
investigated a modified version of the density functional (Eq.~\Ref{eq:df3}) in which the correlation term 
$F^\m{ex,c}_\m{corr}[\set{\rho}]$ is omitted. An asymptotic analysis analogous to Sec.~\ref{sec:asym} leads to
exponentially decaying profiles of $\Delta\set{\rho}$, $\Delta\psi$, and $\rho^Q$. The corresponding phase 
diagram is qualitatively the same as in Fig.~\ref{fig:a}. In particular, there are also first-order wetting 
and drying transition points which are separated by intervals of only partial wetting \emph{and} drying. Due 
to the exponentially decaying electrostatic potential, this modified model is similar to a recently investigated 
model of hard rods interacting with an exponentially decaying wall potential \cite{Shun06}. Whereas the authors 
of Ref. \cite{Shun06} relied entirely on a numerical approach, which provided them only with \emph{evidences} of 
first-order wetting transitions, we shall show analytically that the wetting and drying transitions shown in 
Fig.~\ref{fig:a} are of first order (Secs. \ref{sec:wetting} and \ref{sec:drying}).


\section{\label{sec:wetting}Wetting}

In this section, we discuss the wetting behavior of the model fluid of platelike macroions and salt 
in contact with a charged hard wall. For a wetting scenario, the boundary conditions imposed
on the solutions of the Euler-Lagrange equations (Eq. \Ref{eq:ele2}) require the isotropic bulk 
structure far from the wall. Isotropic-nematic coexistence with such boundary conditions is 
denoted as $\Delta\mu^*_{C_kM} = 0^-$.

In principle, the asymptotic wetting behavior can be completely inferred from the \emph{effective 
interface potential} $\Omega^\m{eff}(\zeta) := (\Omega[\set{\rho}_\zeta]-\Omega_b)/|A|$ 
where $\Omega_b$ denotes the \emph{bulk} contribution to the grand potential and the density profiles 
$\set{\rho}_\zeta$ are the solutions of the Euler-Lagrange equations (Eq. \Ref{eq:ele2}) under the 
\emph{constraint} of a prescribed film thickness $\zeta$ characterized by 
the position of the isotropic-nematic interface \cite{Diet88}. Since exact solutions are out of reach for 
the present model, we have determined $\Omega^\m{eff}(\zeta)$ approximately by considering the following
subspace of density profiles \cite{Diet91}:
\begin{equation}
   \tilde\rho_{i,\zeta}(z) := \left\{
   \begin{array}{ll}
      \rho_i^\m{wn}(z)       & , z \leq \zeta/2 \\
      \rho_i^\m{ni}(z-\zeta) & , z > \zeta/2    \\
   \end{array}
   \right., 
   \label{eq:sharpkink}
\end{equation}
where $\set{\rho}^\m{wn}(z)$ and $\set{\rho}^\m{ni}(z)$ correspond to the wall-nematic and the free 
nematic-isotropic density profiles with the surface and the interface located at $z=0$, respectively. 
Therefore, in the interval 
$z\in(-\infty,\zeta/2]$ the trial density profile $\set{\tilde\rho}_\zeta$ is described by the 
wall-nematic profile whereas in the interval $z\in(\zeta/2,\infty)$ it is given by the free
nematic-isotropic profile shifted to position $\zeta$. Due to $\rho_i^\m{wn}(\infty) = \rho_i^\m{ni}(-\infty)$, 
the discontinuity of $\set{\tilde\rho}_\zeta$ at $z = \zeta/2$ vanishes in the limit $\zeta\rightarrow\infty$.
The transition regime around $z = \zeta/2$ does not contribute to the leading asymptotic terms of 
$\Omega^\m{eff}(\zeta\rightarrow\infty)$ \cite{Diet91}.

Substituting the trial density profiles $\set{\tilde\rho}_\zeta$ into the density functional in Eq. \Ref{eq:df3} 
and using the asymptotic behavior of $\set{\rho}^\m{wn}$ and $\set{\rho}^\m{ni}$ (Sec. \ref{sec:asym}) leads 
to the effective interface potential
\begin{eqnarray}
   \tilde\Omega^\m{eff}(\zeta) 
   & = & 
   \gamma_{wn} + \gamma_{ni} - (\rho_M^\m{ni}(-\infty) - \rho_M^\m{ni}(\infty))\zeta\Delta\mu^*_M
   \nonumber\\
   & &    
   + a_2\zeta^{-2} + a_3\zeta^{-3} + \mathcal{O}(\zeta^{-4})
   \label{eq:omegaeff}
\end{eqnarray}
with the amplitudes $a_2$ and $a_3$ being independent of $\zeta$. Since the density profiles decay proportional to 
$z^{-3}$ towards the bulk values (Sec. \ref{sec:asym}), the results of Ref.~\cite{Diet91} can be used directly:
$a_2$ depends only on the particle charges and bulk densities whereas $a_3$, in addition, contains contributions
due to the wall-nematic and the free nematic-isotropic excess adsorption (see Appendix A). 
The only difference between the density functional in Eq.~\Ref{eq:df3} and the one investigated in 
Ref.~\cite{Diet91} is the presence of the electrostatic term Eq.~\Ref{eq:sumel}. It can be shown that it
merely contributes a term $\mathcal{O}(\zeta^{-4})$ to $\tilde\Omega^\m{eff}(\zeta)$ due to 
the more rapid asymptotic decay of the charge density $\rho^Q(z\rightarrow\infty) = \mathcal{O}(z^{-5})$.

By inspection one recognizes the \emph{same} leading asymptotic decay of $\tilde\Omega^\m{eff}(\zeta)$ 
in Eq.~\Ref{eq:omegaeff} as for systems governed by isotropic \emph{nonretarded dispersion forces} 
\cite{Diet88,Schi90}. We regard this as an surprising result because our model (Sec.~\ref{sec:Fex}) does
\emph{not} include dispersion forces. The asymptotic behavior $\sim z^{-3}$ in Eq.~\Ref{eq:omegaeff} is ultimately 
generated by the assumption of a pair distribution function $g_{ij}$ which decays asymptotically with a 
Debye-H\"{u}ckel (screened Coulomb) form \cite{Bier05}. This assumption concerning the pair distribution 
function $g_{ij}$ is valid because the linearization approximation underlying the Debye-H\"{u}ckel theory is 
justified at large distances.

Minimizing the effective interface potential $\Omega^\m{eff}(\zeta)$ in Eq. \Ref{eq:omegaeff} with 
respect to the interface position $\zeta$ leads to the (equilibrium) excess adsorption 
$\Gamma_M \sim (-\Delta\mu^*_{C_kM})^{-\frac{1}{3}} \nearrow\infty$ for $\Delta\mu^*_{C_kM} \nearrow 0$ as 
long as $a_2 > 0$. Evaluating the analytic expression for $a_2$ along the isotropic-nematic coexistence line, 
which depends only on bulk quantities (see Ref.~\cite{Diet91} and Appendix A), one finds two wetting transition 
points --- corresponding to 
$W^-$ and $W^+$ in Fig.~\ref{fig:a} --- with complete wetting, i.e., $a_2>0$, in between.

\begin{figure}[!t]
   \includegraphics{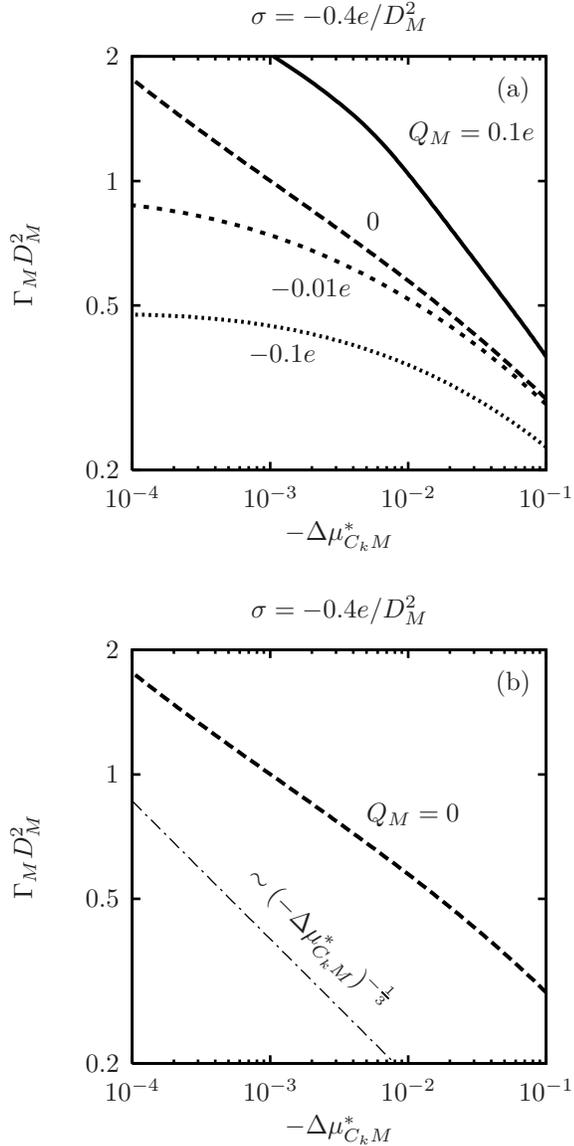}
   \caption{\label{fig:j}Excess adsorption $\Gamma_M$ of platelike macroions along vertical
           thermodynamic paths in Fig. \ref{fig:a} characterized by fixed macroion charges $Q_M$ and 
           parameterized by the chemical potential difference $\Delta\mu^*_{C_kM}< 0$, which measures 
           the thermodynamic distance from isotropic-nematic bulk coexistence. (a) The numerically
           determined excess adsorption $\Gamma_M$ remains finite upon $\Delta\mu^*_{C_kM} \nearrow 0$ 
           for $Q_M \leq -0.01e$ (see also Fig. \ref{fig:b}), whereas a divergence is suggested for the 
           macroion charges $Q_M=0$ and $Q_M=0.1e$, i.e., there is complete wetting of the surface by a 
           nematic film. (b) The comparison of the numerical solution for $Q_M=0$ (dashed line) with the 
           analytical asymptotic power law behavior $\Gamma_M\sim(-\Delta\mu^*_{C_kM})^{-\frac{1}{3}}$  
           (dash-dotted line) (see main text) indicates that the asymptotic regime is not yet reached 
           within the numerically accessible range of undersaturations.}
\end{figure}
Figure~\ref{fig:j} displays the excess adsorption $\Gamma_M$ (Eq. \Ref{eq:GammaM}) as a function of 
the chemical potential difference $\Delta\mu^*_{C_kM} < 0$ for fixed macroion charges $Q_M$. For 
$\Delta\mu^*_{C_kM} \nearrow 0$, isotropic-nematic bulk coexistence is approached along vertical 
thermodynamic paths in Fig.~\ref{fig:a}. In Fig.~\ref{fig:j}(a) the curves for $Q_M=0$ and $Q_M=0.1e$  
suggest a divergence of $\Gamma_M$ in this limit, i.e., \emph{complete wetting}
of the surface by a nematic phase occurs for these macroion charges. On the other hand, for 
$Q_M \leq -0.01e$ the excess adsorption remains finite, i.e., \emph{partial wetting} occurs.
The complete wetting curves for $Q_M=0$ and $Q_M=0.1e$ in Fig.~\ref{fig:j}(a) exhibit \emph{no discontinuity}
in the shown range. Thus, the prewetting lines attached to the first-order wetting transition points $W^-$ and 
$W^+$ must be closer to the isotropic-nematic coexistence line than the numerically accessible values of 
$\Delta\mu^*_{C_kM}$.
 
Figure~\ref{fig:j}(b) compares the analytically obtained asymptotic behavior for the case $Q_M=0$ 
(dash-dotted line) with the corresponding numerical solution (dashed line). The differences between these 
curves indicate that the ultimate asymptotic regime is not yet reached within the numerically accessible 
range of undersaturations.

The order of the wetting transitions at state points $W^-$ and $W^+$ in Fig.~\ref{fig:a} is determined by
$\tilde\Omega^\m{eff}(\zeta)$ for $\Delta\mu^*_{C_kM}=0$. Critical wetting occurs for $a_2=0$, provided 
$a_3>0$; if $a_3<0$ the wetting transition is of first order and does not necessarily occur at the point
given by $a_2=0$ \cite{Diet88,Diet91}. According to Ref. \cite{Diet91} and Appendix A the analytical expression 
for $a_3$
contains a contribution due to the wall-nematic excess adsorption, which is influenced by the surface charge
density $\sigma$. Therefore, there is the possibility that the order of the wetting transition depends on 
the surface charge density $\sigma$. For the values of $\sigma$ used in the present study, however, we 
always find $a_3 < 0$, i.e., the wetting transitions at $W^-$ and $W^+$ are of first order.

Whereas the asymptotical analysis above is reliable with respect to the \emph{order} of the wetting transitions 
at $W^-$ and $W^+$, this is not the case concerning the \emph{location} of $W^-$ and $W^+$, because the wetting 
transitions are of first order \cite{Diet88}. Therefore, we have to use numerical methods.

\begin{figure}[!t]
   \includegraphics{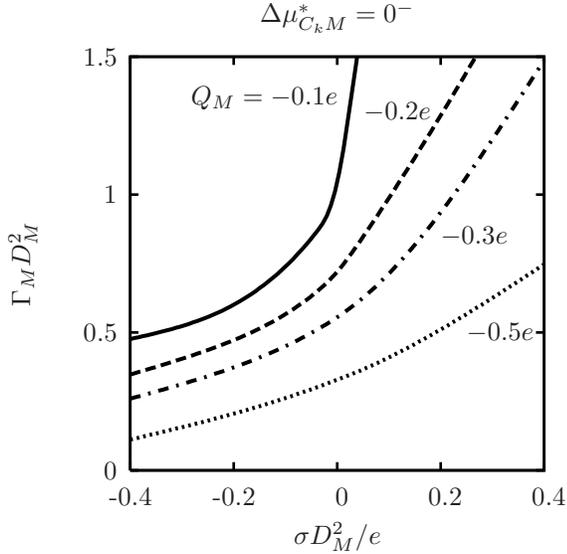}
   \caption{\label{fig:b}Excess adsorption $\Gamma_M$ of platelike macroions at isotropic-nematic 
           coexistence with isotropic boundary conditions in the bulk ($\Delta\mu^*_{C_kM} = 0^-$)
           (see Fig.~\ref{fig:a}) in terms of the surface charge density $\sigma$. Curves for 
           $Q_M = \mean{Q}_M$ and $Q_M = -\mean{Q}_M$ can be mapped upon each other by reflecting them 
           at the axis $\sigma=0$. For fixed macroion charge, the excess adsorption increases 
           with increasing surface charge density. For fixed surface charge density, the excess adsorption 
           \emph{decreases} upon increasing $|Q_M|$, even for attractive walls.}
\end{figure}
In Fig.~\ref{fig:b} the numerically determined excess adsorption at coexistence is shown as a function of the
surface charge density $\sigma$. Without loss of generality, only curves for negative macroion charges 
$Q_M$ are displayed: since the density functional described in Sec. \ref{sec:generalformalism} is invariant 
under the simultaneous inversion of the signs of all charges ($Q_M$, $Q_S$, and $\sigma$), the curve for 
$Q_M=\mean{Q}_M$ is mapped onto the curve for $Q_M = -\mean{Q}_M$ by reflecting it at the axis $\sigma=0$.

As expected, the excess adsorption $\Gamma_M$ of the macroions increases with increasing surface
charge density $\sigma$ for fixed macroion charge $Q_M < 0$, because the surface becomes increasingly
attractive (or decreasingly repulsive) for the macroions. However, for a fixed surface charge density 
$\sigma$ and sufficiently large macroion charges $|Q_M|$, $\Gamma_M$ decreases upon increasing $|Q_M|$, 
irrespective of the sign of $Q_M$, i.e., even for $\sigma Q_M < 0$, for which the wall attracts macroions.
This depression of the macroion number density near the surface occurs because the macroion-macroion 
repulsion dominates the macroion-surface interactions (see Sec. \ref{sec:phasediagram}). From 
Fig.~\ref{fig:b} one can indeed infer that there is \emph{partial} wetting for sufficiently large macroion 
charges $|Q_M|$. 

\begin{figure}[!t]
   \includegraphics{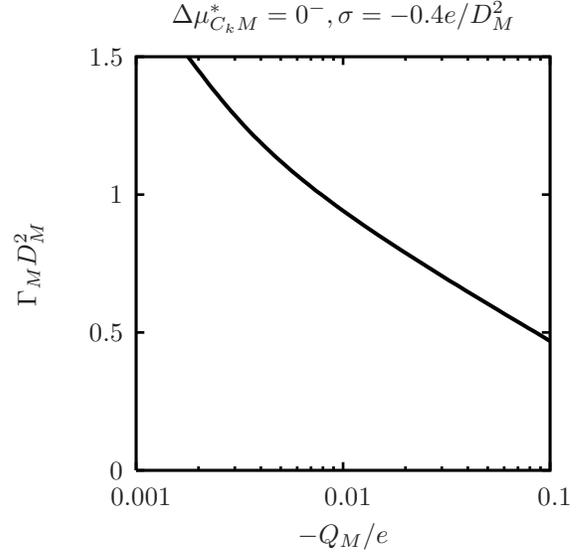}
   \caption{\label{fig:c}Excess adsorption $\Gamma_M$ of platelike macroions at isotropic-nematic 
           coexistence with isotropic boundary conditions in the bulk ($\Delta\mu^*_{C_kM} = 0^-$)
           for a surface charge density $\sigma=-0.4e/D_M^2$ (see Fig.~\ref{fig:a}) as function of the 
           macroion charge $Q_M$. It is finite within the range $-Q_M \geq 2\times 10^{-3}e$ which implies
           the location of the lower wetting transition point $W^-$ (Fig.~\ref{fig:a}) to be within the range 
           $Q_M \in [-2\times 10^{-3}e,0]$.}
\end{figure}
The variation of the macroion excess adsorption $\Gamma_M$ upon $Q_M \nearrow 0$ along
isotropic-nematic coexistence is shown in Fig.~\ref{fig:c}. $\Gamma_M$ is finite for 
$-Q_M \geq 2\times 10^{-3}e$. On the other hand, $Q_M=0$ corresponds to hard colloidal platelets for which
the occurrence of complete wetting is well known \cite{Harn02d}. Hence the lower wetting transition point $W^-$
in Fig.~\ref{fig:a} is located within the range $Q_M^{W^-}\in [-2\times 10^{-3}e,0]$. In order to locate the upper 
wetting transition point $W^+$ one may use the fact that the true wetting transitions points $W^\pm$ can only 
be located \emph{within} the interval $Q_M\in[-Q_M^*,Q_M^*]$, $Q_M^* = 0.1317165(5)e$, where $\pm Q_M^*$ are 
the locations of the wetting transition points inferred from the asymptotic analysis. This statement follows
from the observation that $|Q_M| > Q_M^*$ leads to $a_2 < 0$ which renders $\zeta=\infty$ as a local 
\emph{maximum} of $\Omega^\m{eff}(\zeta)$ \cite{Diet88}. Together with the numerically found complete wetting 
for $Q_M=0.1e$ one concludes that the location of the upper wetting transition point $W^+$ lies within the range 
$Q_M^{W^+}\in [0.1e,0.132e]$.

\begin{figure}[!t]
   \includegraphics{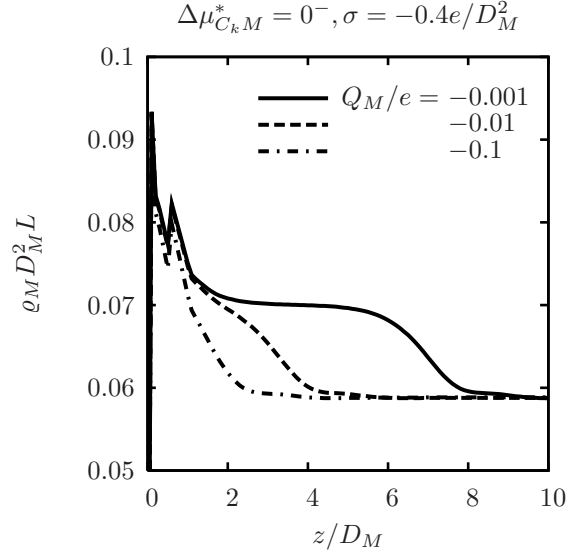}
   \caption{\label{fig:f}Macroion density profiles $\rho_M$ of a mixture of platelike macroions 
           and salt at isotropic-nematic coexistence with isotropic boundary conditions in the bulk
           ($\Delta\mu^*_{C_kM} = 0^-$) for a surface charge density $\sigma=-0.4e/D_M^2$ 
           (see Fig.~\ref{fig:a}). Upon decreasing $|Q_M|$, i.e., upon approaching the wetting
           transition point $W^-$ in Fig.~\ref{fig:a}, a nematiclike film forms at the surface.}
\end{figure}
Finally, Fig.~\ref{fig:f} displays the increase of nematic film thicknesses upon increasing $Q_M < 0$
in terms of the macroion density profiles $\rho_M$.


\section{\label{sec:drying}Drying}

In this section, we study the fluid composed of platelike macroions and salt
in contact with a charged hard surface for nematic boundary conditions
at large distances from the wall. Isotropic-nematic coexistence with nematic boundary 
conditions in the bulk will be denoted as $\Delta\mu^*_{C_kM} = 0^+$.

We have performed an asymptotic analysis of the effective interface potential for drying 
similar to the one described in Sec.~\ref{sec:wetting}. By means of this analysis two first-order 
drying transition points $D^-$ and $D^+$ have been found. Due to the first-order character of the drying 
transitions, the loci of $D^-$ and $D^+$ have been determined numerically (see Sec. \ref{sec:wetting}). Note
that the upper drying transition point $D^+$ is not visible in Fig.~\ref{fig:a} because it is located at 
$Q_M^{D^+}>e$. However, for sufficiently small surface charge densities $|\sigma|$ both drying transition
points have been found within the interval $Q_M\in[-e,e]$. In contrast to the wetting scenario in 
Sec.~\ref{sec:wetting}, we have been able to numerically detect the discontinuity of the excess 
adsorption at the drying transitions. Moreover, the locations of the predrying lines in the
phase diagrams (e.g., Fig.~\ref{fig:a}) can be numerically determined.

\begin{figure}[!t]
   \includegraphics{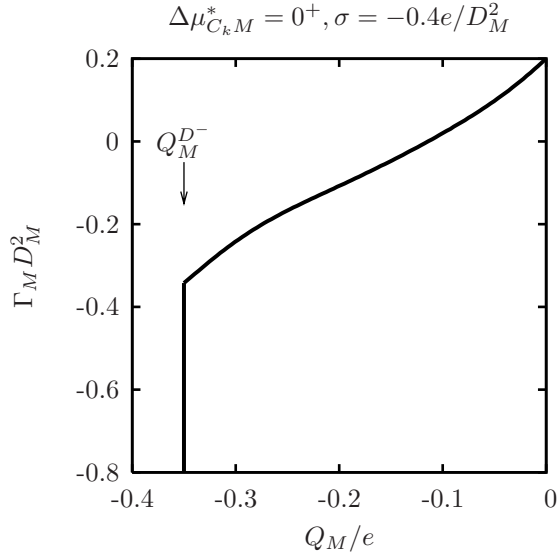}
   \caption{\label{fig:d}Excess adsorption $\Gamma_M$ of platelike macroions at 
           isotropic-nematic coexistence with nematic boundary conditions 
           ($\Delta\mu^*_{C_kM} = 0^+$) for surface charge density $\sigma=-0.4e/D_M^2$
           (see Fig.~\ref{fig:a}) in terms of the macroion charge $Q_M$. The excess adsorption
           is finite and bounded from below for $Q_M > Q_M^{D^-} = -0.35e$, whereas  at $Q_M = Q_M^{D^-}$ it 
           jumps to $-\infty$. It has been verified by a comparison of surface tensions that the 
           numerical solutions with finite excess adsorption correspond to equilibrium structures
           and not only to metastable states. The \emph{discontinuity} of $\Gamma_M$ at $Q_M^{D^-}$ identifies 
           state point $D^-$ in Fig.~\ref{fig:a} as a \emph{first-order} drying transition point. The latter 
           conclusion can also be drawn from an asymptotic analysis of the effective interface potential 
           (see main text).}
\end{figure}
Figure~\ref{fig:d} displays the macroion excess adsorption $\Gamma_M$ close to a charged hard wall
with surface charge density $\sigma = -0.4e/D_M^2$ as a function of the macroion charge $Q_M \leq 0$.
For $-0.35e = Q_M^{D^-} < Q_M \leq 0$, the excess adsorption is finite and bounded from below. It jumps to 
$-\infty$ at $Q_M = Q_M^{D^-}$. The wall-nematic surface tensions of the numerical solutions with finite
excess adsorption are smaller than the sums of the corresponding wall-isotropic surface tensions and the 
isotropic-nematic interfacial tensions. Therefore, these numerical solutions indeed describe equilibrium 
structures and not only metastable states. The \emph{discontinuity} of $\Gamma_M$ corresponds to the 
occurrence of a first-order drying transition at $Q_M^{D^-}$, which is displayed as state point $D^-$ in 
Fig.~\ref{fig:a}.

A first-order drying transition is accompanied by a \emph{predrying line} in the  
surface phase diagram (see the dashed line in Fig.~\ref{fig:a}), which connects the drying transition point
$D^-$ with a critical point $C^-_D$, located at $(Q_M^{C^-_D} = -0.87e, \Delta\mu^{*C^-_D}_{C_kM}=0.049)$. 
The predrying line may be parameterized in terms of, e.g., the macroion charge: 
$\Delta\mu^{*PD}_{C_kM}(Q_M)$ for $Q_M\in[Q_M^{C^-_D},Q_M^{D^-})$ denotes the chemical potential 
difference $\Delta\mu^*_{C_kM}$ for which the excess adsorption $\Gamma_M$ as a function of 
$Q_M$ and $\Delta\mu^*_{C_kM}$ exhibits a \emph{finite} discontinuity $\Delta\Gamma_M$.
\begin{figure}[!t]
   \includegraphics{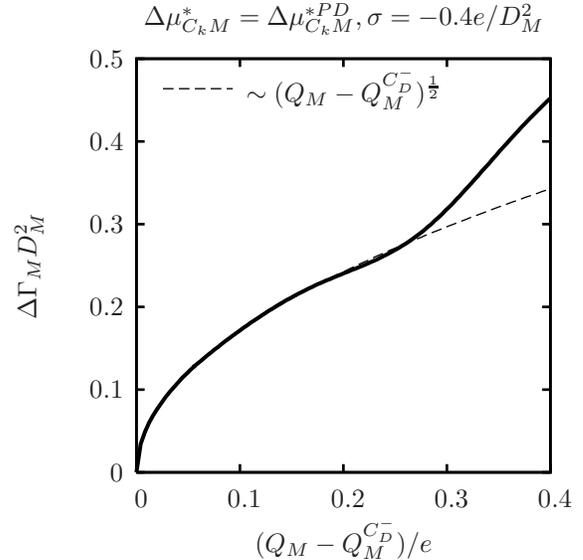}
   \caption{\label{fig:e}Excess adsorption discontinuity $\Delta\Gamma_M$ (solid line)
           of a mixture of platelike macroions and salt at the predrying line parameterized 
           by the macroion charge $Q_M$ for a surface charge density $\sigma=-0.4e/D_M^2$
           (see Fig.~\ref{fig:a}). The excess adsorption difference vanishes for 
           $Q_M \searrow Q_M^{C^-_D} = -0.87e$ according to a power law 
           $\Delta\Gamma_M\sim(Q_M-Q_M^{C^-_D})^{\beta}$ with the mean field critical exponent 
           $\beta = \frac{1}{2}$ (dashed line).}
\end{figure}
Figure~\ref{fig:e} displays this discontinuity as a function of the macroion charge $Q_M$ (solid
line). It vanishes according to a power law $\Delta\Gamma_M \sim (Q_M-Q_M^{C^-_D})^{\beta}$ with
the mean field critical exponent $\beta = \frac{1}{2}$ (dashed line). \emph{Beyond} mean field theory,
one expects an exponent $\beta = \frac{1}{8}$, corresponding to the two-dimensional Ising universality 
class.

\begin{figure}[!t]
   \includegraphics{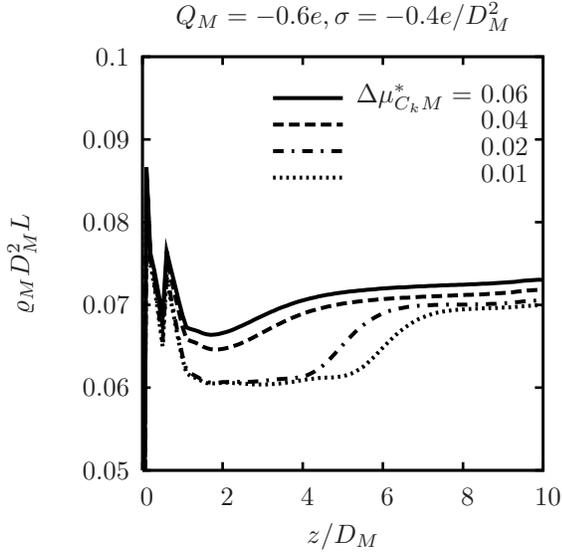}
   \caption{\label{fig:g}Macroion density profiles $\rho_M$ of a mixture of platelike macroions 
           with $Q_M=-0.6e$ and salt in contact with a surface of charge density 
           $\sigma=-0.4e/D_M^2$ upon crossing the predrying line at $\Delta\mu^*_{C_kM}\approx 0.03$ 
           (Fig.~\ref{fig:a}). At the predrying line, a quasi-isotropic film with \emph{finite} 
           thickness appears at the surface.}
\end{figure}
Figure~\ref{fig:g} depicts the formation of an isotropic film upon approaching isotropic-nematic
coexistence for $\Delta\mu^*_{C_kM} > 0$ with macroion charge $Q_M=-0.6e$ and surface charge
density $\sigma=-0.4e/D_M^2$. A \emph{finite} discontinuity of the film thickness upon crossing
the predrying line at $\Delta\mu^*_{C_kM} \approx 0.03$ can be inferred.


\section{\label{sec:electrostaticpotential}Electrostatic potential}

Whereas the three preceeding sections have been focused on the fluid structure, this section addresses the 
electrostatic properties of the surface due to the contact with the fluid of charged particles.

The electrostatic potential profile difference $\Delta\psi(z) := \psi(z) - \psi(\infty)$ (see Eq. 
\Ref{eq:psi3}) relative to the electrostatic bulk potential $\psi(\infty)$ for $Q_M=-0.5e$ at 
isotropic-nematic bulk coexistence with isotropic boundary conditions in the bulk ($\Delta\mu^*_{C_kM}=0^-$)
is shown in 
\begin{figure}[!t]
   \includegraphics{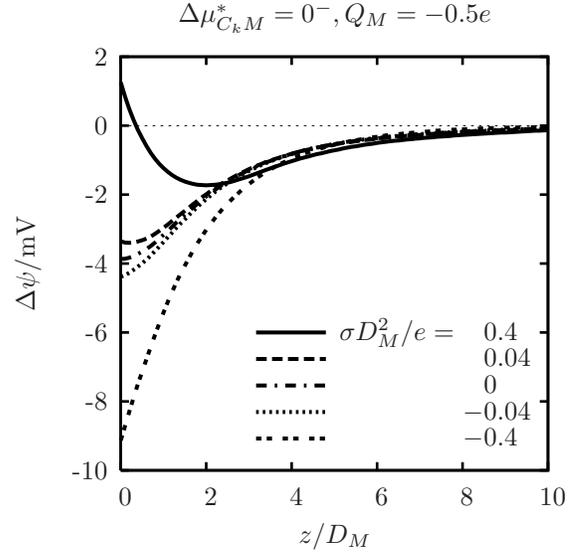}
   \caption{\label{fig:h}Electrostatic potential difference profiles $\Delta\psi$ relative to the 
           electrostatic bulk potential in a mixture of platelike macroions with $Q_M=-0.5e$ and salt at 
           isotropic-nematic bulk coexistence with isotropic boundary conditions in the bulk 
           ($\Delta\mu^*_{C_kM} = 0^-$) (see Fig.~\ref{fig:a}). The slope of $\Delta\psi$ at the surface 
           ($z=0$) is governed by the surface charge density $\sigma$ whereas the bulk value 
           $\Delta\psi(\infty)=0$ is approached proportional to $z^{-3}$ from below. Hence
           upon increasing the surface charge, a crossover from monotonic to non-monotonic
           electrostatic potential profiles occurs.}
\end{figure}
Fig.~\ref{fig:h}. For large distances $z$ from the surface, $\Delta\psi(z)$ decays proportional to $z^{-3}$
(see Sec.~\ref{sec:asym}). Upon increasing the surface charge density $\sigma$ 
a crossover from monotonic to non-monotonic electrostatic potential profiles occurs at $\sigma=0$. According to 
Eq. \Ref{eq:poisson}, the slope of the electrostatic potential difference at the wall is given by the surface 
charge density: $\Delta\psi'(0^+)=-8\sigma$. On the other hand, an electric double-layer is formed with 
a negatively charged layer on the nematic side and a positively charged layer on the isotropic side 
if a quasi-free interface between the isotropic bulk and a nematic film of finite thickness is present
\cite{Bier05}. Thus $\Delta\psi$ approaches its bulk value $0$ from below, i.e., $\Delta\psi'(z)>0$ for large
distances $z$ from the surface. Therefore, for $\sigma>0$, i.e., $\Delta\psi'(0^+)<0$, $\Delta\psi(z)$ is minimal
at some finite distance $0<z_0<\infty$, whereas for $\sigma<0$, i.e., $\Delta\psi'(0^+)>0$, $\Delta\psi$ 
attains its minimal value at $z=0$.

In the aforementioned case, a quasi-free isotropic-nematic interface is formed at isotropic-nematic
coexistence for a state of partial wetting. Additionally, nematic films of finite thickness are
present for states slightly below isotropic-nematic coexistence. Quasi-free isotropic-nematic interfaces 
also occur for nematic boundary conditions in the bulk close to complete drying. In the latter case, the 
electrostatic potential \emph{decreases} upon approaching the bulk because $Q_M<0$. Moreover, we have found 
that the electrostatic potential is a \emph{monotonic} function of the distance from the surface if \emph{no} 
isotropic-nematic interfaces form.

\begin{figure}[!t]
   \includegraphics{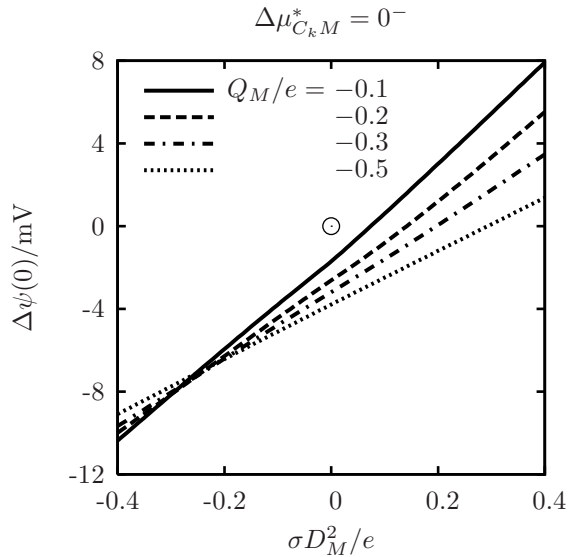}
   \caption{\label{fig:i}Electrostatic surface potential $\Delta\psi(0) = \psi(0)-\psi(\infty)$ as a 
           function of the surface charge density $\sigma$ in a mixture of platelike macroions and salt 
           at isotropic-nematic coexistence with isotropic boundary conditions in the bulk
           ($\Delta\mu^*_{C_kM} = 0^-$) (see Fig.~\ref{fig:a}). All curves are monotonically 
           increasing but they do \emph{not} pass through the origin ($\odot$) at ($\sigma=0, \Delta\psi(0)=0$) 
           if $Q_M\not=0$. Thus, the point of zero (surface) charge ($\sigma=0$) does 
           \emph{not} coincide with the isoelectric point ($\Delta\psi(0)=0$). Note that 
           curves for $Q_M = \mean{Q}_M$ and $Q_M = -\mean{Q}_M$ can be mapped upon each other 
           by reflecting them at the origin.}
\end{figure}
Figure~\ref{fig:i} displays the electrostatic surface potential $\Delta\psi(0)$ as a function
of the surface charge density $\sigma$ at isotropic-nematic coexistence with isotropic boundary
conditions in the bulk ($\Delta\mu^*_{C_kM} = 0^-$). Curves for $Q_M = \mean{Q}_M$ and $Q_M = -\mean{Q}_M$ 
can be mapped upon each other by reflecting them at the origin ($\odot$). As expected, the surface 
potential increases monotonically with the surface charge. However, the points of zero charge 
($\sigma=0$) do \emph{not} coincide with the isoelectric points ($\Delta\psi(0)=0$), i.e., the 
curves in Fig.~\ref{fig:i} do \emph{not} pass through the origin ($\odot$), in contrast to 
Poisson-Boltzmann theories \cite{Grah47}. This behavior arises from the hard-core interaction of the 
particles, which leads to a depletion attraction of the larger macroions towards the wall on purely 
entropic grounds. Therefore, for a hard and uncharged wall ($\sigma=0$), negatively charged macroions ($Q_M<0$)
are accumulated close to the surface leading to a negative electrostatic surface potential ($\Delta\psi(0)<0$).

Our findings show that the sign of the electrostatic surface potential $\Delta\psi(0)$
for vanishing surface charge density $\sigma=0$ depends only on the sign of the macroion charge
and not on the boundary conditions in the bulk or the chemical potential difference $\Delta\mu^*_{C_kM}$.


\section{\label{sec:summary}Summary}

This work has been devoted to the investigation of a recently introduced model
fluid composed of charged platelike colloids and salt (Fig.~\ref{fig:geo}) in
contact with a homogeneously charged planar wall. The density functional introduced
in Ref. \cite{Bier05} has been extended by a wall potential incorporating a 
short-ranged hard-core repulsion as well as a long-ranged Coulomb interaction 
with charged fluid particles. By solving the corresponding Euler-Lagrange 
equations for appropriate boundary conditions, spatially inhomogeneous
equilibrium states are obtained. 

Bulk and surface phase diagrams (Fig.~\ref{fig:a}) in terms of macroion charge, chemical potential 
supersaturation, and surface charge density have been calculated by means of asymptotical analysis of 
effective interface potentials and numerical solutions of the Euler-Lagrange equations. 
The effective interface potentials exhibit the same asymptotic behavior as nonretarded van der Waals
forces although our model does not include dispersion forces. The origin of this phenomenon
can be traced back to the application within the density functional of a particle-particle pair 
distribution function which decays asymptotically with a Debye-H\"{u}ckel form.

Complete wetting by the nematic phase occurs in between two first-order wetting transition points 
on the isotropic-nematic bulk coexistence curve for isotropic boundary conditions in the bulk 
(Fig.~\ref{fig:j}) whereas only partial wetting occurs at the remaining state points of the 
isotropic-nematic coexistence line (Fig.~\ref{fig:b}). The continuous but limited increase of the excess 
adsorption (Fig.~\ref{fig:c}) and of the wetting film thickness (Fig.~\ref{fig:f}) upon approaching the wetting 
transition points is used to locate the wetting transition points within the phase diagrams. Whereas the 
first-order character of the wetting transitions and the complete wetting behavior 
$\Gamma_M \sim (-\Delta\mu^*_{C_kM})^{-\frac{1}{3}}$ have been inferred from analytical considerations, the 
location of the wetting transition points must be obtained numerically. The corresponding prewetting lines 
could not be resolved.

First-order drying by the isotropic phase, characterized by a discontinuous divergence of the excess 
adsorption upon crossing the drying transition point, is found at isotropic-nematic coexistence for nematic 
boundary conditions in the bulk (Fig.~\ref{fig:d}). The predrying line, where the excess adsorption 
shows a finite discontinuity (Fig.~\ref{fig:g}), terminates at a critical point (Fig.~\ref{fig:e}). 

If quasi-free isotropic-nematic interfaces between the bulk and surface films of finite thickness 
form, a crossover is found from monotonic to non-monotonic electrostatic potential profiles
upon varying the surface charge density (Fig.~\ref{fig:h}). Surface 
potential and surface charge do \emph{not} vanish simultaneously (Fig.~\ref{fig:i}),
i.e., the point of zero charge and the isoelectric point of the surface do not coincide
due to the presence of both Coulomb interactions and hard-core repulsion. 


\appendix\section{}

The $5\times5$-matrix $A_{ij}$ in Eq.~\Ref{eq:deltach} is given by
\begin{equation}
   A_{ij} = -\int\limits_{-\infty}^\infty\!\mathrm{d}z\;c^\m{h}_{ij}\big(z,[\set{\rho}(\infty)]\big),
\end{equation}
where $c^\m{h}_{ij}$ denotes the two-particle direct correlation function of the hard core
term $F^\m{ex,h}$ (Eq.~\Ref{eq:Fexh}) and $\set{\rho}(\infty)$ is the set of \emph{constant} 
bulk density profiles.

By defining
\begin{eqnarray}
   w_{ij}(z) & := &\Int{A(L)}{2}{\vec{a}} U^\m{c}_{ij}(\vec{a},z)\big(\exp(-U^\m{h}_{ij}(\vec{a},z))-1
   \nonumber\\
   & & 
   + \exp(-U^\m{h}_{ij}(\vec{a},z)) G_{ij}(\kappa,\snorm{\vec{a},z})
   \big),
\end{eqnarray}
which renders Eq.~\Ref{eq:Fexccorr} in the form
\begin{equation}
   F^\m{ex,c}_\m{corr}[\set{\rho}] 
   =
   \frac{1}{2}|A|\sum_{ij}\int\limits_{0^-_{\vphantom{|}}}^\infty\!\mathrm{d}z
   \int\limits_{0^-_{\vphantom{|}}}^\infty\!\mathrm{d}z'
   \rho_i(z)\rho_j(z')w_{ij}(z-z'),
\end{equation}
the $5 \times 5$-matrix $B_{ij}$ in Eq.~\Ref{eq:deltacccorr} can be expressed as
\begin{equation}
   B_{ij} = \int\limits_{-\infty}^\infty\!\mathrm{d}z\;w_{ij}(z).
\end{equation}

According to Ref.~\cite{Diet91}, the amplitudes $a_2$ and $a_3$ in Eq.~\Ref{eq:omegaeff} are given by
\begin{equation}
   a_2 = \frac{1}{2}\sum_{ij}T_{ij}
\end{equation}
and
\begin{equation}
   a_3 = \sum_{ij}T_{ij}(d_{i,\m{wn}} - d_{j,\m{ni}})
\end{equation}
with the abbreviation
\begin{equation}
   T_{ij} := -\frac{24Q_i^2Q_j^2}{\kappa^4}\rho_i^\m{ni}(-\infty)
             \big(\rho_j^\m{ni}(-\infty)-\rho_j^\m{ni}(\infty)\big).
\end{equation}
The quantities
\begin{equation}
   d_{i,\m{wn}} := \int\limits_0^\infty\!\mathrm{d}z\Bigg(1 - \frac{\rho_i^\m{wn}(z)}{\rho_i^\m{wn}(\infty)}\Bigg)
\end{equation}
and
\begin{eqnarray}
   d_{j,\m{ni}} & := & \frac{1}{\rho_j^\m{ni}(-\infty)-\rho_j^\m{ni}(\infty)}\times
   \nonumber\\
   & & \Bigg(\int\limits_{-\infty}^0\!\mathrm{d}z\big(\rho_j^\m{ni}(z)-\rho_j^\m{ni}(-\infty)\big) + 
   \nonumber\\
   & &
   \phantom{\Bigg(}
   \int\limits_0^\infty\!\mathrm{d}z\big(\rho_j^\m{ni}(z)-\rho_j^\m{ni}(\infty)\big)\Bigg)
\end{eqnarray}
are related to excess adsorptions of the wall-nematic surface and the free nematic-isotropic interface,
respectively.




\begin{thebibliography}{00}
\bibitem{Units}   Dimensionless quantities are used in units of $\beta^{-1}=k_BT$ (thermal
                     energy), $e$ (proton charge), and $\ell_B=\frac{\beta e^2}{4\pi\epsilon_0\epsilon}$
                     (Bjerrum length) with the solvent dielectric constant $\epsilon$. For 
                     $T=300\,\m{K}$ and $\epsilon=78$ (water) the latter is $\ell_B \approx 
                     0.72\,\m{nm}$. In order to avoid a clumsy notation, the corresponding dimensional
                     quantities are denoted by the same symbol and can be recognized by the presence 
                     of units. Dimensional number densities are sometimes given in units of 
                     $1\,\m{mM} = 1\,\m{mol/m^3} \approx 6 \times 10^{23}\,\m{m^{-3}}$.
\end{thebibliography}
\end{document}